# Comparison of different versions of SignalP and TargetP for diatom plastid protein predictions with ASAFind

Ansgar Gruber, Cedar McKay, Gabrielle Rocap, Miroslav Oborník

Biology Centre, Institute of Parasitology, Czech Academy of Sciences; School of Oceanography, University of Washington; Biology Centre, Institute of Parasitology, Czech Academy of Sciences, Faculty of Science, University of South Bohemia

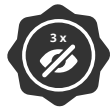

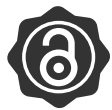

**Full Open Access**
Supported by the Velux Foundation, the University of Zurich, and the EPFL School of Life Sciences.

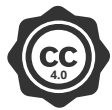

## Abstract

Plastid targeted proteins of diatoms and related algae can be predicted with high sensitivity and specificity using the ASAFind method published in 2015. ASAFind predictions rely on SignalP predictions of endoplasmic reticulum (ER) targeting signal peptides. Recently (in 2019), a new version of SignalP was released, SignalP 5.0. We tested the ability of SignalP 5.0 to recognize signal peptides of nucleus-encoded, plastid-targeted diatom pre-proteins, and to identify the signal peptide cleavage site. The results were compared to manual predictions of the characteristic cleavage site motif, and to previous versions of SignalP. SignalP 5.0 is less sensitive than the previous versions of SignalP in this specific task, and also in the detection of signal peptides of non-plastid proteins in diatoms. However, in combination with ASAFind, the resulting prediction performance for plastid proteins is high. In addition, we tested the multi-location prediction tool TargetP for its suitability to provide signal peptide information to ASAFind. The newest version, TargetP 2.0, had the highest prediction performances for diatom signal peptides and mitochondrial transit peptides compared to other versions of SignalP and TargetP, thus it provides a good basis for ASAFind predictions.

## Introduction

Diatoms are important primary producers in many aquatic habitats [3] and are increasingly important for biotechnological applications [4] [5]. Genome and transcriptome sequencing has revealed unusual aspects of diatom metabolism, many of which are related to the intracellular location of metabolic pathways in the diatom cell [6]. Predicting intracellular locations of proteins from sequence data is therefore an important step in diatom genome annotation [7] [8] [9].

Plastid proteins in diatoms are either plastid-encoded or are nucleus-encoded and need to be targeted to the complex diatom plastids, across four envelope membranes [10] [11]. Nucleus encoded plastid proteins of diatoms are recognized by characteristic bipartite targeting signals, consisting of a signal peptide domain, a transit peptide domain, a conserved motif at the signal peptide cleavage site, and another conserved motif at the transit peptide cleavage site [12] [13] [14] [15] [10] [16] [17] [18] [19] [20] [21].

Plastid proteomes of two species of diatoms have been predicted with the dedicated prediction program ASAFind, and new sequence data can be analyzed with the ASAFind web service or a downloadable software [1]. ASAFind has also been used to predict plastid proteins of the cryptophyte *Guillardia theta* [22], and other organisms with complex plastids of red algal origin [23] [24] [25]. ASAFind builds upon the prediction of ER signal peptides by SignalP, a software that was released in different versions, which employ different prediction algorithms and therefore also differ in their predictions and performance [26] [27] [28].

## Objective

Since the release of the ASAFind prediction method in 2015 [1], new versions of SignalP [29] and TargetP [30] have been released. The objective of this study is to find out whether these updated software tools improve the prediction of diatom plastid targeting pre-sequences with ASAFind, in comparison to the versions of SignalP with which ASAFind was originally designed to work.

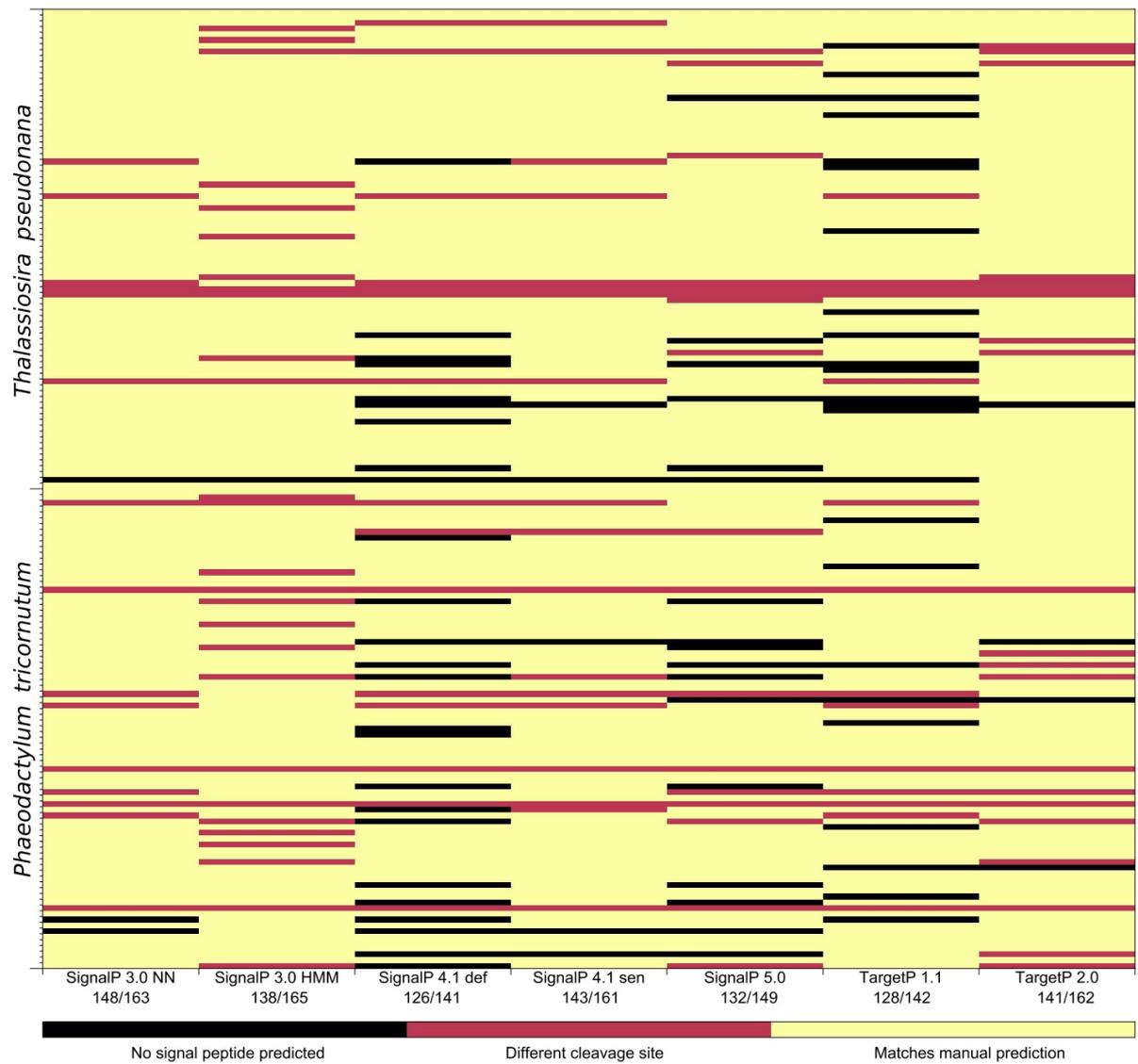

a

**Signal peptide prediction**

| Prediction method | sensitivity | specificity | MCC |
|---|---|---|---|
| TargetP 2.0 | 0.84 | 0.94 | 0.60 |
| SignalP 4.1 sensitive | 0.87 | 0.82 | 0.56 |
| SignalP 5.0 | 0.79 | 0.94 | 0.53 |
| SignalP 3.0 HMM | 0.92 | 0.65 | 0.53 |
| TargetP 1.1 | 0.83 | 0.82 | 0.51 |
| SignalP 4.0 | 0.77 | 0.88 | 0.48 |
| SignalP 4.1 default | 0.77 | 0.88 | 0.48 |
| SignalP 3.0 NN | 0.90 | 0.59 | 0.45 |

**Plastid protein prediction**

| Prediction method | ASAFind confidence | sensitivity | specificity | MCC |
|---|---|---|---|---|
| SignalP 4.1 sensitive + ASAFind | high only | 0.87 | 0.99 | 0.88 |
| TargetP 2.0 + ASAFind | high only | 0.85 | 0.99 | 0.86 |
| SignalP 5.0 + ASAFind | high only | 0.84 | 0.99 | 0.85 |
| TargetP 2.0 + ASAFind | low or high | 0.96 | 0.88 | 0.84 |
| SignalP 3.0 NN + ASAFind | high only | 0.80 | 0.99 | 0.82 |
| SignalP 4.1 sensitive + ASAFind | low or high | 0.98 | 0.84 | 0.81 |
| SignalP 4.0 + ASAFind | high only | 0.78 | 0.99 | 0.80 |
| SignalP 4.1 default + ASAFind | high only | 0.78 | 0.99 | 0.80 |
| SignalP 5.0 + ASAFind | low or high | 0.91 | 0.90 | 0.80 |
| TargetP 1.1 + ASAFind | high only | 0.75 | 0.97 | 0.76 |
| SignalP 4.0 + ASAFind | low or high | 0.89 | 0.87 | 0.75 |
| SignalP 4.1 default + ASAFind | low or high | 0.89 | 0.87 | 0.75 |
| SignalP 3.0 NN + ASAFind | low or high | 0.93 | 0.82 | 0.74 |
| TargetP 1.1 + ASAFind | low or high | 0.85 | 0.86 | 0.71 |
| HECTAR | not applicable | 0.71 | 0.94 | 0.67 |
| SignalP 3.0 HMM + ASAFind | high only | 0.60 | 0.97 | 0.64 |
| SignalP 3.0 HMM + ASAFind | low or high | 0.73 | 0.84 | 0.58 |

**Mitochondrial transit peptide prediction**

| Prediction method | sensitivity | specificity | MCC |
|---|---|---|---|
| TargetP 2.0 | 0.70 | 0.98 | 0.68 |
| TargetP 1.1 | 0.70 | 0.88 | 0.41 |

b

# Figure Legend

**Figure 1A.**

Comparison of results from prediction servers with manually predicted signal peptide cleavage site positions in the training set of 83 orthologous pairs of putative plastid-targeted protein sequences from *Thalassiosira pseudonana* and *Phaeodactylum tricornutum* used to generate the ASAFind scoring matrix (found in Table S1 of [1], raw data from the prediction servers are attached). SignalP 3.0 uses either a neural network (NN) or a hidden Markov (HMM) model. SignalP 4.1 was tested with the default (def) and with the sensitive (sen) options, SignalP 5.0 employs a deep neural network, see methods for references. Sequences are plotted in the same order as they appear in the raw data files. The numbers below each prediction server indicate the number of cleavage sites that matched the manual prediction/ and the total number of signal peptides predicted.

**Figure 1B.**

Sensitivities, specificities and Matthews correlation coefficients (MCC) of the different variations of SignalP and TargetP for the prediction of signal peptides, mitochondrial transit peptides and plastid proteins (in conjunction with ASAFind), determined with a set of 132 experimentally localized reference proteins from multiple intracellular destinations in *Phaeodactylum tricornutum* (found in Table S3 of [1]). Results are sorted from highest to lowest MCC. The performance of HECTAR [2] is also listed for comparison.

## Results & Discussion

The ASAFind plastid protein prediction [1] is based on the prediction of signal peptides by the prediction software SignalP [26] [28]. Predictions of signal peptides, as well as positions of predicted cleavage sites in diatom plastid targeting pre-sequences, differ between the SignalP versions available in 2007 [18]. Because ASAFind performs a sliding window motif search surrounding the position of the SignalP predicted signal peptide cleavage site, the choice of the most suitable version of SignalP is crucial for the performance of ASAFind [1]. We tested the training set of 166 putative plastid-targeted protein sequences used to build the scoring matrix for the ASAFind method with the newly released SignalP 5.0 [29] and with two versions of TargetP [31] [32] [30], with which ASAFind has not been tested. SignalP 5.0 predicted fewer signal peptides than most previous versions of SignalP (149 of the 166 signal peptides). TargetP 1.1 predicted only 142 signal peptides while TargetP 2.0 predicted 162 signal peptides (see Fig. 1A and corresponding raw data).

In addition to the prediction of a signal peptide, SignalP and TargetP also predict the position of the signal peptide cleavage site [26]. ASAFind uses the predicted position of the cleavage site to determine the range of the sequence windows in which the sequence is scored according to the conserved motif [1]. This is in part because there is considerable variation in the predicted positions of the signal peptide cleavage sites between SignalP 3.0 NN, HMM, and SignalP 4.1, the versions of SignalP that were available in 2015 [1]. SignalP 5.0, TargetP 1.1, and TargetP 2.0 also differ in the prediction of the cleavage site position between sequences (see Fig. 1A and corresponding raw data). Notably, while there are some sequences where the predicted signal peptide cleavage site is different from the manually predicted one across several versions of the prediction programs, there are many other examples of individual sequences for which different versions of the prediction programs give different results.

Due to the sliding window approach of ASAFind, the exact position of the SignalP predicted cleavage site is less important for the prediction result than the actual score of the highest-scoring sequence window [1]. We therefore also tested the performance of ASAFind in combination with SignalP 5.0, TargetP 1.1, and TargetP 2.0 using the reference set of 132 experimentally localized proteins from various intracellular locations in *Phaeodactylum tricornutum* (Table S3 of [1]). Sensitivities, specificities, and MCCs are compiled in figure 1B (raw data also attached). For plastid proteins, the highest MCC is obtained by combining ASAFind with the SignalP 4.1 "sensitive" setting (Fig. 1B). For overall signal peptide predictions (of plastid as well as non-plastid proteins), and also for mitochondrial transit peptide predictions, TargetP 2.0 has the best performance by MCC (Fig. 1B). Because TargetP 2.0 also performs well specifically for plastid proteins in combination with ASAFind (Fig. 1B), TargetP 2.0 is a valuable tool for protein targeting analyses in diatoms.

## Conclusions

SignalP 5.0 is a highly specific prediction program for signal peptides and differentiates between three types of signal peptides in Bacteria and Archaea. While its overall performance in the organisms for which it was developed is higher than that of its predecessors [29], our results show that it is at the same time, less sensitive in the detection of signal peptides in proteins from diatoms. Nevertheless, it gives good results for diatom plastid protein prediction in conjunction with ASAFind. TargetP 2.0 [30] has the highest Matthews correlation coefficients for the prediction of signal peptides and mitochondrial transit peptides in diatoms and thus is also suitable as a basis for ASAFind plastid protein predictions.

## Limitations

The 132 protein reference set of [1] was originally compiled with the goal to evaluate the performance of plastid protein predictions (55 positives, 77 negatives, which contained all experimental protein localizations in *Phaeodactylum tricornutum* published at the time). It is highly asymmetric with respect to other sequence categories. For mitochondria, it only contains 10 positive sequences (versus 122 negatives), and for signal peptides, it only contains 17 negative sequences (versus 115 positive sequences, including the 55 plastid-targeted proteins). These ratios can lead to over-optimistic performance measure in the case of sensitivity and specificity. We therefore also included MCCs, which account for asymmetries in the reference set. The MCCs are low for the signal peptide and mitochondrial transit peptide predictions tested with this dataset, in comparison to the MCCs for plastid protein predictions (Fig. 1B).

## Alternative Explanations

The prediction performances of SignalP and TargetP for signal peptides and mitochondrial transit peptides of diatoms might actually be higher than what the low MCCs (Fig. 1B) suggest, if they could be tested with more symmetric reference datasets (see "Limitations").

## Conjectures

In the future, larger reference sets will certainly help to improve the characteristics of the prediction methods. Additional studies with experimental protein localizations in diatoms have been published since 2015 (e.g. [33] [34] [35] [36] [37] [38]), and should be included in future reference sets. Here, we used the reference set from [1], in order to keep the results comparable to the previously published study.

## Additional Information

### Methods

The sequences used to calculate the ASAFind scoring matrices (training set, Table S1 from [1]) and the reference protein set (Table S3 from [1]) were downloaded from the journal homepage. For these sequences, prediction results of SignalP 3.0 NN [39], SignalP 3.0 HMM [40], SignalP 4.1 [41] [27] (with default and "sensitive" settings) and SignalP 5.0 [29] (with the "Eukarya" organism group selected) were obtained from the SignalP web server (http://www.cbs.dtu.dk/services/SignalP/) (if not already contained in Table S1 from [1]). TargetP results were obtained from the TargetP web servers (http://www.cbs.dtu.dk/services/TargetP/) for TargetP 1.1 [31] [32] (with the default "no cutoffs; winner-takes-all" option, and the "non-plant" organism group selected) and TargetP 2.0 [30] ("non-plant" organism group selected). In both TargetP variants, "non-plant" should be used as the organism group, because otherwise, TargetP will also search for chloroplast transit peptides known from primary plastids of higher plants and green algae, a type of targeting pre-sequence that does not exist in diatoms. The results were re-formatted to ASAFind input requirements with spreadsheet software and a text editor. For SignalP 5.0 and TargetP 2.0, the content of the "Prediction" and "CS Position" columns of the output table was used to start ASAFind scoring of the sequences with predicted signal peptides; for TargetP 1.1, the content of the "Loc" and "TPlen" columns was used (with "TPlen" corrected by +1 to obtain the cleavage site position). The collected results were analyzed using spreadsheet software and custom Perl scripts. The performance statistics of HECTAR [2] on the reference set was reproduced from [1]. For the calculation of the prediction performance of signal peptides, all sequences from the reference set that enter the secretory pathway were counted as "positive", all others were counted as "negative"; for the calculation of the prediction performance of plastid proteins, all experimentally plastid localized proteins from the reference set were counted as "positive", all others were counted as "negative"; for the calculation of the prediction performance of mitochondrial proteins, all experimentally mitochondria localized proteins from the reference set were counted as "positive", all others were counted as "negative" (see raw data file for the exact assignments). Sensitivities, specificities, and Matthews correlation coefficients (MCC) were calculated according to the formulas given in [42].

## Funding Statement

Our research is supported by the Institute of Parasitology of the Czech Academy of Sciences, by the Grant Agency of the Czech Republic (18-13458S), by the ERDF/ESF Centre for research of pathogenicity and virulence of parasites (No.CZ.02.1.01/0.0/0.0/16_019/0000759) and by the US National Science Foundation (G.R and C.M).

## Ethics Statement

Not Applicable.

# Citations